\documentclass[cccite]{cc}
\usepackage{times}
\usepackage{amsfonts}
\usepackage{amssymb}
\usepackage{amsmath}
\usepackage{graphicx}
\usepackage{amsthm}

\newcommand{\ket}[1]{|#1\rangle}

\newcommand{\bra}[1]{\langle #1|}

\newcommand{\E}{{\rm E}}
\newcommand{\SSS}{\mathcal{S}}

\newcommand{\tr}{\mbox{Tr}}
\newcommand{\diag}{\mbox{diag}}

\def\half{\textstyle{\frac{1}{2}}}
\def\quarter{\textstyle{\frac{1}{4}}}
\def\01{\{0,1\}}

\def\ket#1{\mbox{$| #1 \rangle$}}

\newtheorem{thm}{Theorem}
\newtheorem{prop}[thm]{Proposition}
\newtheorem{lem}[thm]{Lemma}
\newtheorem{cor}[thm]{Corollary}

\title{Perfect Parallel Repetition Theorem for Quantum XOR Proof Systems}

\titlehead{Perfect Parallel Repetition for Quantum XOR Proof Systems}

\contact{cleve@cs.uwaterloo.ca}
\received{\today}
\author{Richard Cleve\\
School of Computer Science and Institute for Quantum Computing\\
University of Waterloo\\
Perimeter Institute for Theoretical Physics\\
Waterloo, Ontario, Canada
\and
William Slofstra\\
Department of Mathematics\\
University of California at Berkeley\\
Berkeley, California, USA\\
Research carried out while at the University of Waterloo
\and
Falk Unger\\
CWI\\
Amsterdam, The Netherlands\\
\and
Sarvagya Upadhyay\\
School of Computer Science and Institute for Quantum Computing\\
University of Waterloo\\
Waterloo, Ontario, Canada}


\begin{subject}
  81P68, 68Q10
\end{subject}

\begin{abstract}
We consider a class of two-prover interactive proof systems where
each prover returns a single bit to the verifier and the verifier's
verdict is a function of the XOR of the two bits received.
We show that, when the provers are allowed to coordinate
their behavior using a shared entangled quantum state, a perfect
parallel repetition theorem holds in the following sense.
The prover's optimal success probability for simultaneously playing
a collection of XOR proof systems is exactly the product
of the individual optimal success probabilities.
This property is remarkable in view of the fact that, in the classical
case (where the provers can only utilize classical information),
it does not hold.
The theorem is proved by analyzing parities of XOR proof systems
using semidefinite programming techniques, which we then relate to
parallel repetitions of XOR games via Fourier analysis.
\end{abstract}
\begin{document}

\maketitle


\section{Introduction and summary of results}

\noindent
The theory of interactive proof systems has played an important role in
the development of computational complexity and cryptography.
Also, the impact of quantum information on the theory of interactive proof
systems has been shown to have interesting consequences
\citep{Watrous99,KitaevW00}.
In \citep{Ben-OrG+88} a variant of the model of interactive proof system
was introduced where there are two provers who have unlimited computational
power subject
to the condition that they cannot communicate between themselves once
the execution of the protocol starts.
This model is sufficiently powerful to characterize NEXP \cite{BabaiF+91}.

Our present focus is on \textit{XOR interactive proof systems},
which are based on \textit{XOR games}.
For a predicate $f : S \times T \rightarrow \{0,1\}$ and a probability
distribution $\pi$ on $S \times T$, define the XOR game $G = (f,\pi)$
operationally as follows.
\begin{itemize}
\item
The Verifier selects a pair of questions $(s,t) \in S \times T$ according
to distribution $\pi$.
\item
The Verifier sends one question to each prover: $s$ to prover Alice and $t$
to prover Bob (who are forbidden from communicating with each other once
the game starts).
\item
Each prover sends a bit back to the Verifier: $a$ from Alice and $b$ from
Bob.
\item
The Verifier accepts if and only if $a \oplus b = f(s,t)$.
\end{itemize}
A definition that is essentially equivalent to this%
\footnote{Except that \textit{degeneracies} are allowed, where for some
$(s,t)$ pairs, the Verifier is allowed to accept or reject independently
of the value of $a \oplus b$. All results quoted here apply to nondegenerate
games.}
appears in \cite{CleveH+04a}.
In the classical version, the provers have unlimited computing power, but
are restricted to possessing classical information; in the quantum version,
the provers may possess qubits whose joint state is entangled.
In both versions, the communication between the provers and the
verifier is classical.

An \textit{XOR interactive proof system} (with soundness probability $s$ and
completeness probability $c > s$) for a language $L$ associates an XOR game with
every input string $x$, such that:
\begin{itemize}
\item
$S_x$ and $T_x$ consist of strings of length polynomial in $|x|$,
$\pi_x$ can be sampled in time polynomial in $|x|$, and $f_x$ can be computed
in time polynomial in $|x|$.
\item
If $x \in L$ then the maximum acceptance probability over prover's strategies
is at least~$c$.
\item
If $x \not\in L$ then the maximum acceptance probability over prover's strategies
is at most~$s$.
\end{itemize}

In \cite{CleveH+04a} it is pointed out that results in \cite{BellareG+98,Hastad01}
imply that, in the case of classical provers, these proof systems have sufficient
expressive power to recognize every language in NEXP (with soundness probability
$s = 11/16 + \epsilon$ and completeness probability $c = 12/16 - \epsilon$, for
arbitrarily small $\epsilon > 0$).
Thus, although these proof systems appear restrictive, they can recognize
any language that an unrestricted multi-prover interactive proof system
can.
Moreover, in \cite{CleveH+04b,Wehner06} it is shown that any language
recognized by a quantum XOR proof system is in EXP.
Thus, assuming EXP $\neq$ NEXP, quantum entanglement strictly weakens
the expressive power of XOR proof systems.

Returning to XOR games, quantum physicists have, in a sense, been
studying them since the 1960s, when John Bell introduced his celebrated
results that are now known as Bell inequality violations \cite{Bell64}.
An example is the \textit{CHSH} game, named after the authors of
\cite{ClauserH+69}.
In this game,
$S = T = \{0,1\}$, $\pi$ is the unform distribution on $S \times T$,
and $f(s,t) = s \wedge t$.
It is well known that, for the \textit{CHSH} game, the best possible
classical strategy succeeds with probability $3/4$, whereas the best
possible quantum strategy succeeds with higher probability
%
of $(1+1/\sqrt2)/2 \approx 0.85$ \cite{ClauserH+69,Tsirelson80}.


Following \cite{CleveH+04a}, for an XOR game $G$, define its
\textit{classical value} $\omega_c(G)$ as the maximum possible success
probability achievable by a classical strategy.
Similarly, define its \textit{quantum value} $\omega_q(G)$ as the
maximum possible success probability achievable by a quantum
strategy.

\subsection{Taking the sum of XOR games}\label{subsec:additivity}

For any two XOR games $G_1 = (f_1, \pi_1)$ and $G_2 = (f_2, \pi_2)$,
define their \textit{sum (modulo $2$)} as the XOR~game
\begin{equation}
G_1 \oplus G_2 = (f_1 \oplus f_2, \pi_1 \times \pi_2).
\end{equation}
In this game, the verifier begins by choosing questions
 $((s_1,t_1),(s_2,t_2)) \in (S_1 \times T_1) \times (S_2 \times T_2)$
according to the product distribution $\pi_1 \times \pi_2$, sending $(s_1,s_2)$
to Alice and $(t_1,t_2)$ to Bob.
Alice and Bob then win if and only if their respective outputs, $a$ and $b$,
satisfy $a \oplus b = f_1(s_1,t_1) \oplus f_2(s_2,t_2)$.

A simple way for Alice and Bob (who may or may not share entanglement) to play
$G_1 \oplus G_2$ is to optimally play $G_1$ and $G_2$ separately, producing
outputs $a_1$, $b_1$ for $G_1$ and $a_2$, $b_2$ for $G_2$, and then to output
$a = a_1 \oplus a_2$ and $b = b_1 \oplus b_2$ respectively.
It is straightforward to calculate that
the above method for playing $G_1 \oplus G_2$ succeeds with
probability
\begin{eqnarray}\label{eq:parity-lb}
\omega(G_1)\omega(G_2) + (1-\omega(G_1))(1-\omega(G_2)).
\end{eqnarray}
Is this the optimal way to play $G_1 \oplus G_2$?

The answer is \textit{no} for \textit{classical} strategies.
To see why this is so, note that, using this approach for
the XOR game $\text{\it CHSH} \oplus \text{\it CHSH}$, produces
a success probability of $5/8$.
A better strategy is for Alice to output $a = s_1 \wedge s_2$ and
Bob to output $b = t_1 \wedge t_2$.
It is straightforward to verify that this latter strategy succeeds
with probability $3/4$.

Our first result is that the answer is \textit{yes} for \textit{quantum}
strategies.
\begin{thm}[\textbf{Additivity}]\label{thm:additivity}
For any two XOR games $G_1$ and $G_2$ an optimal quantum strategy
for playing $G_1 \oplus G_2$ is for Alice and Bob to optimally
play $G_1$ and $G_2$ separately, producing outputs $a_1$, $b_1$
for $G_1$ and $a_2$, $b_2$ for $G_2$, and then to output
$a = a_1 \oplus a_2$ and $b = b_1 \oplus b_2$.
\end{thm}

The proof of Theorem~\ref{thm:additivity} uses a known characterization
of quantum strategies for individual XOR games as semidefinite programs.
Section~\ref{sec:additivity} contains the proof.

\subsection{Parallel repetition of XOR games}\label{subsec:parallel}

For any sequence of XOR games $G_1 \! = \! (f_1,\pi_1),\, \dots$,
$G_n = (f_n,\pi_n)$,
define their \textit{conjunction}, denoted by $\wedge_{j=1}^n G_j$, as follows.
The verifier chooses questions
$((s_1,t_1),\dots,(s_n,t_n))
\in (S_1 \times T_1) \times \dots \times (S_n \times T_n)$ according to the
product distribution $\pi_1 \times \dots \times \pi_n$, and sends
$(s_1,\dots,s_n)$ to Alice and $(t_1,\dots,t_n)$ to Bob.
Alice and Bob output bits $a_1,\dots,a_n$ and $b_1,\dots,b_n$, respectively,
and win if and only if their outputs simultaneously satisfy these $n$
conditions:
$a_1 \oplus b_1 = f_1(s_1,t_1)$, $\dots,$ $a_n \oplus b_n = f_n(s_n,t_n)$.
(Note that $\wedge_{j=1}^n G_j$ is not itself an XOR game for $n > 1$.)

One way for Alice and Bob to play $\wedge_{j=1}^n G_j$ is to
independently play each game optimally.
This succeeds with probability $\prod_{j=1}^n\omega(G_j)$.
Is this the optimal way to play $\wedge_{j=1}^n G_j$?

The answer is \textit{no} for classical strategies.
It is shown in \cite{BarrettC+02} that%
\footnote{After posing this question about
$\omega_c(\mbox{\textit{CHSH}} \wedge \mbox{\textit{CHSH}})$,
the answer was first shown to us by S.~Aaronson,
who independently discovered the classical protocol and then
found the prior result in~\cite{BarrettC+02}.}
$\omega_c(\mbox{\textit{CHSH}} \wedge \mbox{\textit{CHSH}})
= 10/16 > 9/16 = \omega_c(\mbox{\textit{CHSH}})\omega_c(\mbox{\textit{CHSH}})$.

Our second result is that the answer is \textit{yes} for quantum strategies.
\begin{thm}[\textbf{Parallel Repetition}]\label{thm:parallel}
For any XOR games $G_1, \dots, G_n$, we have that
$\omega_q(\wedge_{j=1}^n G_j) = \prod_{j=1}^n\omega_q(G_j)$.
\end{thm}
This is a quantum version of Raz's parallel repetition theorem~\cite{Raz98}
for the restricted class of XOR games.
We call it a \textit{perfect} parallel repetition theorem because the
probabilities are multiplicative in the exact sense (as opposed to an
asymptotic sense, as in~\cite{Raz98}).
The proof of Theorem~\ref{thm:parallel} is based on Theorem~\ref{thm:additivity}
combined with Fourier analysis techniques for boolean functions.
Section~\ref{sec:parallel} contains the proof.

\subsection{Comparison with other work}

There is no known parallel repetition theorem along the lines of \cite{Raz98}
for quantum games (where the players share entanglement).
As far as we know, our results represent the first progress in this direction.
Recently, Holenstein~\cite{Holenstein06} gave a simplified proof of the parallel
repetition theorem that applies to classical and no-signalling strategies.
Neither of these cases capture quantum strategies for XOR games (for example,
every XOR game has value 1 in the no-signaling model).

For games other than XOR games, the question of parallel repetition remains
open.
Watrous~\cite{Watrous04} has shown that, there is a binary game (that is not
an XOR game) for which $\omega_q(G) = \omega_q(G \wedge G) = 2/3$, as in the
classical case.
For completeness, this is shown in Appendix~A.
This implies that a perfect parallel repetition property does not automatically
apply to quantum games.


For a broad class of games, Feige and Lov\'asz~\cite{FeigeL92} define
quantities that are relaxations---and hence upper bounds---of their classical
values, and show that one of these quantities satisfies a parallel repetition
property analogous to Theorem~\ref{thm:parallel}.
For any XOR game $G$, the Feige-Lov\'asz relaxations of its classical
value are equal to the quantum value of $G$.
Although this was also noted previously~\cite{FeigeK+07,FeigeG95}, for
completeness, an explicit proof of this is shown in Appendix~B.
It is important to note that, \textit{for general games}, the relationship
between their quantum values and the Feige-Lov\'asz relaxations of their
classical values are not understood.
As far as we know, neither quantity bounds the other for general games.
However, using the fact that they are equivalent for XOR games combined with
our Theorem~\ref{thm:parallel}, we deduce (in Appendix B) that, whenever
$G_1, \dots, G_n$ are XOR games, the quantum value of $\wedge_{j=1}^n G_j$
coincides with its associated Feige-Lov\'asz relaxations.
(Note that this does not reduce our Theorem~\ref{thm:parallel} to the results
in~\cite{FeigeL92}, since we invoke Theorem~\ref{thm:parallel} to deduce that
the quantum value and the Feige-Lov\'asz relaxations are the same for
$\wedge_{j=1}^n G_j$.)


\section{Proof of the Additivity Theorem}\label{sec:additivity}

In this section we prove Theorem~\ref{thm:additivity}, which is stated in
Section~\ref{subsec:additivity}.

It is convenient to define the quantum \textit{bias} of an XOR game
as $\varepsilon_q(G) = 2\omega_q(G) - 1$.
Then, due to Eq.~\ref{eq:parity-lb}, to prove Theorem~\ref{thm:additivity},
it suffices to show that
$\varepsilon_q(G_1 \oplus G_2) = \varepsilon_q(G_1) \varepsilon_q(G_2)$.

Since Alice and Bob can independently play games $G_1$ and $G_2$
optimally and then take the parity of their outputs as their outputs for
$G_1 \oplus G_2$, we immediately have the following.
\begin{prop}\label{prop:lb}
For two XOR games $G_1$ and $G_2$,
$\varepsilon_q(G_1 \oplus G_2) \ge \varepsilon_q(G_1) \varepsilon_q(G_2)$.
\end{prop}
\noindent The nontrivial part of the proof is the reverse inequality.

A quantum strategy for an XOR game consists of a bipartite quantum state
$\ket{ \psi}$ shared by Alice and Bob, a set of observables $X_{s}$ $(s \in S)$
corresponding to Alice's part of the quantum state, and a set of observables
$Y_{t}$ $(t \in T)$ corresponding to Bob's part of the state.
The bias achieved by this strategy is given by
\begin{equation*}
\sum_{s,t} \pi(s,t)(-1)^{f(s,t)}
\left\langle \psi \lvert X_{s} \otimes Y_{t} \rvert \psi \right\rangle.
\end{equation*}

We make use of a vector characterization of XOR games due to \cite{Tsirelson:85b}
(also pointed out in \cite{CleveH+04a}), which is a consequence of the following.
\begin{thm}[\textbf{\cite{Tsirelson:85b,CleveH+04a}}]\label{Tsirel}
Let S and T be finite sets, and let $\ket{\psi}$ be a
pure quantum state with support on a bipartite Hilbert space
$\cal{H} = \cal{A} \otimes \cal{B}$ such that
$\mbox{dim}(\cal{A}) = \mbox{dim}(\cal{B}) = $ n.
For each $s \in S$ and $t \in T$, let $X_{s}$ and $Y_{t}$ be observables
on $\cal{A}$ and $\cal{B}$ with eigenvalues $\pm 1$ respectively.
Then there exists real unit vectors $x_{s}$ and $y_{t}$ in
$\mathbb{R}^{2n^{2}}$ such that
\begin{equation*}
\left\langle \psi \lvert X_{s} \otimes Y_{t} \rvert \psi \right\rangle =
x_{s} \cdot y_{t},
\end{equation*}
for all $s \in S$ and $t \in T$.
\\
Conversely, suppose that S and T are finite sets, and $x_{s}$ and $y_{t}$ are
unit vectors in $\mathbb{R}^{N}$ for each $s \in S$ and $t \in T$.
Let $\cal{A}$ and $\cal{B}$ be Hilbert space of dimension
$2^{\lceil N/2 \rceil}$, $\cal{H} = \cal{A} \otimes \cal{B}$ and
$\ket{\psi}$ be a maximally entangled state on $\cal{H}$.
Then there exists observables
$X_{s}$ and $Y_{t}$ with eigenvalues $\pm 1$, on $\cal{A} \text{ and } \cal{B}$
respectively, such that
\begin{equation*}
\left\langle \psi \lvert X_{s} \otimes Y_{t} \rvert \psi \right\rangle =
x_{s} \cdot y_{t},
\end{equation*}
for all $s \in S$ and $t \in T$.
\end{thm}

Using Theorem~\ref{Tsirel}, we can characterize Alice and Bob's quantum
strategies by a choice of unit vectors $\{x_s\}_{s \in S}$ and
$\{y_t\}_{t \in T}$.
Using this characterization, the bias becomes
\begin{equation}\label{E:bias}
\varepsilon_q(G) = \max_{\{x_{s}\},\{y_{t}\}}
\quad \sum_{s,t} \pi(s,t)(-1)^{f(s,t)}~x_{s} \cdot y_{t}.
\end{equation}
The {\it cost matrix} for the XOR game $G$ is defined as the matrix $A$ with entries
$A_{s,t} = \pi(s,t) (-1)^{f(s,t)}$.

Note that any matrix $A$, with the provision that the absolute values of the
entries sum to~1, is the cost matrix of an XOR game.
If $G_1$ and $G_2$ are XOR games with cost matrices $A_1$ and $A_2$ respectively,
then the cost matrix of $G_1 \oplus G_2$ is $A_1 \otimes A_2$.
Also, for $0 \leq \lambda \leq 1$, define the convex combination
$\lambda G_1 + (1-\lambda) G_2$ to be the XOR game with cost matrix
\begin{equation*}
    \left( \begin{array}{cc} 0  & \lambda A_1 \\
                           (1-\lambda) A_2 & 0
            \end{array} \right).
\end{equation*}
This convex combination can be interpreted as the game where, with
probability $\lambda$, game $G_1$ is played and, with probability
$1 - \lambda$, game $G_2$ is played (and Alice and Bob are informed
about which game is occurring).
Also, for a game $G$ with cost matrix $A$, define $G^T$ to be the game
with cost matrix $A^T$.
In other words, Alice and Bob switch places to play $G^T$.
The next proposition summarizes some simple facts.
\begin{prop}\label{P:simplefacts} \
\newline
$1.$
$\varepsilon_q(G_1 \oplus G_2) = \varepsilon_q(G_2 \oplus G_1)$ and
$\varepsilon_q(G) = \varepsilon_q(G^T)$.
\vspace*{1.0mm} \newline
$2.$
For all $ 0 \le \lambda \le 1$, \vspace*{0.5mm} \newline
$\varepsilon_q\left(\lambda G_1 + \left(1-\lambda \right) G_2 \right)
= \lambda \varepsilon_q(G_1) + \left( 1 - \lambda \right) \varepsilon_q(G_2)$
\vspace*{0.5mm} \newline
\hspace*{3mm} and \vspace*{0.5mm} \newline
{$G_1 \!\oplus \left(\lambda  G_2 + (1-\lambda)G_3 \right)
= \lambda (G_1 \!\oplus G_2) + (1-\lambda)(G_1 \!\oplus G_3)$.}
\end{prop}

The bias of a quantum XOR game may be stated as a semidefinite programming
problem (SDP).
We refer to Boyd and Vandenberghe \cite{Boyd:04} for a detailed introduction
to semidefinite programming.
For cost matrix $A$, the bias is equivalent to the objective value of problem
\begin{equation*}\label{E:intermedprob}
\max~\tr \left(A^T U_{1}^{T}U_{2}\right) : \diag \left(U_1^T U_1\right)
= \diag\left(U_2^T U_2\right) = \bar{e},
\end{equation*}
where $\{x_s\}$ and $\{y_t\}$ appear as the columns of $U_1$ and $U_2$ respectively.
Here $\diag(M)$ denotes the column vector of diagonal entries of the matrix $M$, and
$\bar{e}$ is the column vector  $(1, \dots ,1)^{T}$.
We begin by considering the game $\half G + \half G^T$, whose
cost matrix
\begin{eqnarray*}
B =
\left( \begin{array}{cc} 0 & \frac{1}{2}A \\
\frac{1}{2}A^T & 0 \end{array} \right)
\end{eqnarray*}
has useful structural properties, one of them being that it is symmetric.
Proposition~\ref{P:simplefacts} implies that
$\varepsilon_q(\half G + \half G^T) = \varepsilon_q(G)$.
This enables us to express the value of game $G$ in terms of the SDP
$(\text{P}_{B})$ defined by
\begin{equation*}
\max~\tr(B X) \quad : \quad \diag(X) = \bar{e}, \quad X \succeq 0.
\end{equation*}
The notation $X \succeq Y$ means that the matrix $X - Y$ lies in the cone of positive
semidefinite matrices.
That $(\text{P}_B)$ is equivalent to problem (\ref{E:intermedprob}) follows from
the fact that a semidefinite matrix $X$ can be written as
$(U_{1}, U_{2})^{T} (U_{1}, U_{2})$ for some matrices $U_1$ and $U_2$.

To show that an optimal solution for $(\text{P}_B)$ exists, we can examine the
Lagrange-Slater dual of $(\text{P}_B)$.
The dual, denoted by $(\text{D}_B)$, is defined as
\begin{equation*}
\min~ (x,y)\bar{e} \quad : \quad \Delta(x,y) \succeq B,
\end{equation*}
where $\Delta(x,y)$ denotes the diagonal matrix with entries given by the (row)
vectors $x,y$.
Both $(\text{P}_B)$ and $(\text{D}_B)$ have Slater points---that is, feasible
points in the interior of the semidefinite cone.
Explicitly, the identity matrix is a Slater point for $(\text{P}_B)$, and
$\bar{e}$ is a Slater point for $(\text{D}_B)$.
Therefore, by the strong duality theorem,
the optimal values of $(\text{P}_B)$ and $(\text{D}_B)$ are the same
and both problems have optimal solutions attaining this value.

The next lemma establishes the upper bound for the game
$(\half G_1 + \half G_1^T) \oplus (\half G_2 + \half G_2^T)$ (which we will show
afterwards has the same bias as $G_1 \oplus G_2$).
\begin{lem}\label{L:dual}
If $G_1$ and $G_2$ are XOR games, then \vspace*{1mm} \newline
$\varepsilon_q((\half G_1 + \half G_1^T) \oplus (\half G_2 + \half G_2^T))
\leq \varepsilon_q(G_1) \varepsilon_q(G_2)$.
\end{lem}
\begin{proof}
Let $G_1$ and $G_2$ be two games with cost matrices $A_1$ and $A_2$,
respectively, and let
\begin{eqnarray}\label{eq:B1andB2}
B_1 =
\left(\!\!\begin{array}{cc}0 & \frac{1}{2}A_1 \\
\frac{1}{2}A_1^T & 0\end{array}\!\!\right)
\ \text{and} \
B_2 =
\left(\!\!\begin{array}{cc}0 & \frac{1}{2}A_2 \\
\frac{1}{2}A_2^T & 0\end{array}\!\!\right).
\end{eqnarray}
Let $(x_1,y_1)$ and $(x_2,y_2)$ be optimal solutions to $(\text{D}_{B_1})$
and $(\text{D}_{B_2})$, respectively, which implies
$\Delta(x_i,y_i) - B_i \succeq 0$ and $\varepsilon_q(G_i)=(x_i,y_i)\bar e$, for $i=1,2$.
It suffices to show that $(x_1,y_1)\otimes(x_2,y_2)$ is a solution to $(\text{D}_{B_1 \otimes B_2})$, since $B_1\otimes B_2$ is
the cost matrix of $(\half G_1 + \half G_1^T) \oplus (\half G_2 + \half G_2^T)$.
Note that, for \textit{arbitrary} $B_1$ and $B_2$,
$\Delta(x_1,y_1) \succeq B_1$ and $\Delta(x_2,y_2) \succeq B_2$
does \textit{not} imply that
$\Delta(x_1,y_1) \otimes \Delta(x_2,y_2) \succeq B_1 \otimes B_2$
(a simple counterexample is when $\Delta(x_1,y_1) = \Delta(x_2,y_2) = 0$
and $B_1 = B_2 = -I$).
We make use of the structure of $B_1$ and $B_2$ arising from
Eq.~\ref{eq:B1andB2}.
For each $i$, $\Delta(x_i,y_i) - B_i \succeq 0$ implies that, for all
(row) vectors $u,v$,
\begin{eqnarray*}
0 \, & \le & \
{\left(\!\begin{array}{cc}u & v\end{array}\!\right) \atop }\!
\left(\!\begin{array}{cc}\Delta(x_i) & -\frac{1}{2}A_i \\
-\frac{1}{2}A_i^T & \Delta(y_i)\end{array}\!\right)\!
\left(\!\begin{array}{c}u^T \\ v^T \end{array}\!\right) \\
& = &
{\left(\!\begin{array}{cc}u & \!\! -v\end{array}\!\right) \atop }\!
\left(\!\begin{array}{cc}\Delta(x_i) & +\frac{1}{2}A_i \\
+\frac{1}{2}A_i^T & \Delta(y_i)\end{array}\!\right)\!
\left(\!\!\!\begin{array}{r}u^T \\ -v^T \end{array}\!\right),
\end{eqnarray*}
which in turn implies that $\Delta\left(x_i,y_i\right) + B_i\succeq 0$
also holds.
Therefore,
\begin{eqnarray*}
\left(\Delta(x_1,y_1) - B_1\right)\otimes \left(\Delta(x_2,y_2) + B_2\right)
& \succeq & 0 \quad \text{and} \nonumber \\
\left(\Delta(x_1,y_1) + B_1\right)\otimes \left(\Delta(x_2,y_2) - B_2\right)
& \succeq & 0,
\end{eqnarray*}
which, by averaging, yields
\begin{equation*}
\Delta(x_1,y_1) \otimes \Delta(x_2,y_2) - B_1 \otimes B_2 \succeq 0.
\end{equation*}

Therefore, $(x_1,y_1) \otimes (x_2,y_2)$ is a feasible point in the dual
$(\text{D}_{B_1 \otimes B_2})$, which obtains the objective value
$\varepsilon_q(G_1) \varepsilon_q(G_2)$, which implies the Lemma.
\end{proof}

Now we may complete the proof of Theorem~\ref{thm:additivity}.
Using Proposition~\ref{prop:lb} for line (\ref{in1}), Lemma~\ref{L:dual}
for line (\ref{in2}) and Proposition~\ref{P:simplefacts} and some easy
algebra for the rest we can derive the following
\begin{eqnarray*}
\lefteqn{\varepsilon_q\left(G_1 \oplus G_2\right)} & & \\
\label{in1}&\ge& \varepsilon_q\left(G_1\right)\varepsilon_q\left(G_2\right)\\
\label{in2}&\ge& \varepsilon_q((\half G_1 + \half G_1^T)
\oplus (\half G_2 + \half G_2^T)) \\
&=& \varepsilon_q\left(\quarter (G_1 \oplus G_2)
+ \quarter (G_1 \oplus G_2^T)
+ \quarter (G_1^T \oplus G_2)
+ \quarter (G_1^T \oplus G_2^T)\right)\\
&=& \varepsilon_q\left(\half \left[\half (G_1 \oplus G_2)
+ \half (G_1 \oplus G_2^T)\right]
+ \half \left[\half (G_1 \oplus G_2)
+ \half (G_1 \oplus G_2^T)\right]^T\,\right)\\
&=& \half \varepsilon_q\left(G_1 \oplus G_2\right)
+ \half \varepsilon_q\left(G_1 \oplus G_2^T\right).
\end{eqnarray*}
Therefore
$\varepsilon_q(G_1 \oplus G_2)  \ge \varepsilon_q(G_1 \oplus G_2^T)$.
By symmetry,
$\varepsilon_q(G_1 \oplus G_2^T) \ge \varepsilon_q(G_1 \oplus G_2)$,
as well, which means that all of the above inequalities must be equalities.
This completes the proof of Theorem~\ref{thm:additivity}.


\section{Parallel repetition theorem}\label{sec:parallel}

In this section we prove Theorem~\ref{thm:parallel}, which is stated in
Section~\ref{subsec:parallel}.

We begin with the following simple probabilistic lemma.

\begin{lem}\label{lemma:bias}
For any sequence of binary random variables $X_1, X_2, \dots, X_n$,
\begin{equation*}
\frac{1}{2^n}\sum_{M \subseteq [n]} \E\left[ (-1)^{\oplus_{j \in M} X_j}\right]
=
\Pr[X_1 \dots X_n = 0 \dots 0].
\end{equation*}
\end{lem}
\begin{proof}
By the linearity of expectation,
\begin{eqnarray*}
\lefteqn{\frac{1}{2^n}\sum_{M \subseteq [n]}
\E\left[ (-1)^{\oplus_{j \in M} X_j}\right]} & & \\
& = & \E\biggl[\frac{1}{2^n}\sum_{M \subseteq [n]}(-1)^{\oplus_{j \in M} X_j}\biggr]
\\
& = & \E\biggl[\ \, \prod_{j=1}^n\left(\frac{1+(-1)^{X_j}}{2}\right)\ \biggr] \\
& = & \Pr\left[X_1 \dots X_n = 0 \dots 0\right],
\end{eqnarray*}
where the last equality follows from the fact that
\begin{equation*}
\prod_{j=1}^n(1+(-1)^{X_j}) \neq 0
\end{equation*}
only if $X_1 \dots X_n = 0 \dots 0$.
\end{proof}

We introduce the following terminology.
For any strategy $\SSS$ (classical or quantum) for any game $G$, define
$\omega(\SSS,G)$ as the success probability of strategy $\SSS$ on game $G$.
Similarly, define the corresponding bias as
$\varepsilon(\SSS,G) = 2\omega(\SSS,G) - 1$.

Now let $\SSS$ be any protocol for the game $\wedge_{j=1}^n G_j$.
For each $M \subseteq [n]$, define the protocol $\SSS_M$
(for the game $\oplus_{j \in M}G_j$) as follows.
\begin{enumerate}
\item
Run protocol $\SSS$, yielding $a_1,\dots,a_n$ for Alice and $b_1,\dots,b_n$ for Bob.
\item
Alice outputs $\oplus_{j\in M} a_j$ and Bob outputs $\oplus_{j\in M} b_j$.
\end{enumerate}

\begin{lem}
\begin{equation*}
\frac{1}{2^n}\sum_{M \subseteq [n]}\varepsilon(\SSS_M,\oplus_{j \in M}G_j)
=
\omega(\SSS,\wedge_{j=1}^n G_j).
\end{equation*}
\end{lem}
\begin{proof}
For all $j \in [n]$, define $X_j = a_j \oplus b_j \oplus f_j(s_j,t_j)$.
Then, for all $M \subseteq [n]$, we have $\E[(-1)^{\oplus_{j \in M} X_j}] =
\varepsilon(\SSS_M,\oplus_{j \in M} G_j)$,
and $\Pr[X_1 \dots X_n = 0 \dots 0] = \omega(\SSS,\wedge_{j=1}^n G_j)$.
The result now follows from Lemma~\ref{lemma:bias}.
\end{proof}

\begin{cor}
\begin{equation}\label{eqn:c-parallel-additivity}
\omega_c(\wedge_{j=1}^n G_j) \le
\frac{1}{2^n} \sum_{M \subseteq [n]} \varepsilon_c(\oplus_{j \in M}G_j)
\end{equation}
and
\begin{equation}\label{eqn:q-parallel-additivity}
\omega_q(\wedge_{j=1}^n G_j) \le
\frac{1}{2^n} \sum_{M \subseteq [n]} \varepsilon_q(\oplus_{j \in M}G_j).
\end{equation}
\end{cor}

Now, to complete the proof of Theorem~\ref{thm:parallel},
using Theorem~\ref{thm:additivity}, we have
\begin{eqnarray}
\frac{1}{2^n} \sum_{M \subseteq [n]} \varepsilon_q(\oplus_{j \in M}G_j)
& = &
\frac{1}{2^n} \sum_{M \subseteq [n]} \prod_{j \in M}\varepsilon_q(G_j)
\nonumber \\
& = & \prod_{j=1}^n\left(\frac{1+\varepsilon_q(G_j)}{2}\right)
\nonumber \\
& = & \prod_{j=1}^n \omega_q(G_j).
\end{eqnarray}
Combining this with Eq.~\ref{eqn:q-parallel-additivity}, we deduce
$\omega_q(\wedge_{j=1}^n G_j) = \prod_{j=1}^n \omega_q(G_j)$,
which completes the proof of Theorem~\ref{thm:parallel}.

\vspace*{3mm}

\noindent\textbf{Comments:} Although Eq.~\ref{eqn:q-parallel-additivity} is
used to prove a tight upper bound on $\omega_q(\wedge_{j=1}^n G_j)$,
Eq.~\ref{eqn:c-parallel-additivity} cannot be used to obtain a tight
upper bound on $\omega_c(\wedge_{j=1}^n G_j)$ for general XOR games.
This is because
$\varepsilon_c(\mbox{\textit{CHSH}})
= \varepsilon_c(\mbox{\textit{CHSH}} \oplus \mbox{\textit{CHSH}}) = 1/2$
and it can be shown that
$\varepsilon_c(\mbox{\textit{CHSH}} \oplus \mbox{\textit{CHSH}}
\oplus \mbox{\textit{CHSH}}) = 5/16$.
Therefore, for $G_1 = G_2 = G_3 = \mbox{\textit{CHSH}}$, the right side
of Eq.~\ref{eqn:c-parallel-additivity} is
$\frac{1}{8} \sum_{M \subseteq [3]} \varepsilon_c(\oplus_{j \in M}G_j)
= 34.5/64$, whereas $\omega_c(\wedge_{j=1}^3 G_j)$ must be expressible as
an integer divided by 64 (in fact%
\footnote{This was independently calculated by S. Aaronson and B. Toner,
by searching over a finite number of deterministic classical strategies.},
$\omega_c(\wedge_{j=1}^3 G_j) = 31/64$).


\begin{acknowledge}
We would like to thank Scott Aaronson, Ben Toner, John Watrous, and Ronald
de Wolf for helpful discussions.
RC and SU acknowledge support from Canada's NSERC, CIAR and MITACS, and the
U.S. ARO/DTO.
WS acknowledges support from Canada's NSERC.
FU acknowledges support from the EU project QAP (IST-2005-15848).
\end{acknowledge}

%

\bibliographystyle{plain}

\begin{thebibliography}{10}




\bibitem{BabaiF+91}
L.~Babai, L.~Fortnow, and C.~Lund.
\newblock Non-deterministic exponential time has two-prover interactive
  protocols.
\newblock {\em Computational Complexity}, 1(1):3--40, 1991.

\bibitem{BarrettC+02}
J.~Barrett, D.~Collins, L.~Hardy, A.~Kent, and S.~Popescu.
\newblock Quantum nonlocality, Bell inequalities and the memory loophole.
\newblock {\em Physical Review A} 66:042111, 2002.

\bibitem{Bell64}
J.~Bell.
\newblock On the {Einstein-Podolsky-Rosen} paradox.
\newblock {\em Physics}, 1(3):195--200, 1964.

\bibitem{BellareG+98}
M.~Bellare, O.~Goldreich, and M.~Sudan.
\newblock Free bits, \uppercase{PCP}s, and non-approximability --- towards
 tight results.
\newblock {\em SIAM Journal on Computing}, 27(3):804--915, 1998.

\bibitem{Ben-OrG+88}
M.~Ben-Or, S.~Goldwasser, J.~Kilian, and A.~Wigderson.
\newblock Multi-prover interactive proofs: how to remove intractability
  assumptions.
\newblock In {\em Proceedings of the Twentieth Annual ACM Symposium on Theory
  of Computing}, pages 113--131, 1988.

\bibitem{Boyd:04}
S.~Boyd and L.~Vandenberghe.
\newblock Semidefinite programming.
\newblock {\em SIAM Review}, 38(1):49--95, 1996.






\bibitem{ClauserH+69}
J.~F. Clauser, M.~A. Horne, A.~Shimony, and R.~A. Holt.
\newblock Proposed experiment to test local hidden-variable theories.
\newblock {\em Physical Review Letters}, 23(15):880--884, 1969.

\bibitem{CleveH+04a}
R.~Cleve, P.~H\o yer, B.~Toner, J.~Watrous.
\newblock Consequences and limits of nonlocal strategies.
\newblock In {\em Proceedings of the 19th IEEE Conference on Computational
Complexity}, pages 236--249, 2004.

\bibitem{CleveH+04b}
R.~Cleve, P.~H\o yer, B.~Toner, and J.~Watrous.
\newblock Consequences and limits of nonlocal strategies.
\newblock Presentation given at {\em 19th IEEE Conference on
Computational Complexity}, June 2004.



\bibitem{Feige91}
U.~Feige.
\newblock On the success probability of two provers in one-round proof systems.
\newblock In {\em Proceedings of the Sixth Annual Conference on Structure in
  Complexity Theory}, pages 116--123, 1991.

\bibitem{FeigeG95}
U.~Feige and M.~Goemans.
\newblock Approximating the value of two prover proof systems, with applications
\newblock to MAX 2SAT and MAX DICUT.
\newblock In {\em Proceedings of the Third Israel Symposium on Theory of
Computing and Systems}, pages 182--189, 1995.

\bibitem{FeigeK+07}
U.~Feige, G.~Kindler, and R.~O'Donnell.
\newblock Understanding parallel repetition requires understanding foams.
\newblock In {\em Proceedings of the 21st IEEE Conference on
Computational Complexity}, pages 179--192, 2007.

\bibitem{FeigeL92}
U.~Feige and L.~Lov\'asz.
\newblock Two-prover one-round proof systems: their power and their problems.
\newblock In {\em Proceedings of the Twenty-Fourth Annual ACM Symposium on
  Theory of Computing}, pages 733--744, 1992.



\bibitem{Fortnow89}
L.~Fortnow.
\newblock Complexity theoretic aspects of interactive proof systems.
\newblock PhD thesis, Massachusetts Institute of Technology, May 1989.
\newblock Technical Report MIT/LCS/TR-447.

\bibitem{FortnowR+94}
L.~Fortnow, J.~Rompel, and M.~Sipser.
\newblock On the power of multi-prover interactive protocols.
\newblock {\em Theoretical Computer Science}, 134:545--557, 1994.





\bibitem{Hastad01}
J.~H{\aa}stad.
\newblock Some optimal inapproximability results.
\newblock {\em Journal of the ACM}, 48(4):798--859, 2001.


\bibitem{Holenstein06}
T.~Holenstein.
\newblock Parallel repetition: simplifications and the no-signaling case.
\newblock In {\em Proceedings of the Thirty-Ninth Annual ACM Symposium on
  Theory of Computing}, pages 411--419, 2007.



\bibitem{KitaevW00}
A.~Kitaev and J.~Watrous.
\newblock Parallelization, amplification, and exponential time simulation of
\newblock quantum interactive proof systems.
\newblock In {\em Proceedings of the Thirty-Second Annual ACM Symposium on
  Theory of Computing}, pages 608--617, 2000.

\bibitem{KobayashiM03}
H.~Kobayashi and K.~Matsumoto.
\newblock Quantum multi-prover interactive proof systems with limited
\newblock prior entanglement.
\newblock {\em Journal of Computer and System Sciences}, 66(3):429--450, 2003.








\bibitem{Raz98}
R.~Raz.
\newblock A parallel repetition theorem.
\newblock {\em SIAM Journal on Computing}, 27(3):763--803, 1998.


\bibitem{Tsirelson80}
B.~S. {(Tsirelson) Cirel'son}.
\newblock Quantum generalizations of {Bell's} inequality.
\newblock {\em Letters in Mathematical Physics}, 4(2):93--100, 1980.

\bibitem{Tsirelson:85b}
B.~S. {(Tsirelson) Tsirel'son}.
\newblock Quantum analogues of the \uppercase{B}ell inequalities:
  \uppercase{T}he case of two spatially separated domains.
\newblock {\em Journal of Soviet Mathematics}, 36:557--570, 1987.


\bibitem{Watrous99}
J.~Watrous.
\newblock PSPACE has constant-round quantum interactive proof systems.
\newblock in {\em Proceedings of the Fourtieth Annual Symposium on Foundations
of Computer Science}, pages 112--119, 1999.

\bibitem{Watrous04}
J.~Watrous.
\newblock Personal communication, 2004.

\bibitem{Wehner06}
S.~Wehner.
\newblock Entanglement in interactive proof systems with binary answers.
\newblock In {\em Proceedings of STACS 2006}, pages 162--171, 2006.

\end{thebibliography}

\section*{Appendix A}

In this appendix, we give the unpublished proof due to Watrous~\cite{Watrous04}
that there is a binary game $G$ (that is not an XOR game) for which
$\omega_q(G) = \omega_q(G \wedge G) = 2/3$.
The game used was originally proposed by Fortnow, Feige and
Lov\'{a}sz~\cite{Fortnow89,FeigeL92}, who showed that
$\omega_c(G) = \omega_c(G \wedge G) = 2/3$.

The game has binary questions ($S = T = \{0,1\}$) and binary answers
($A = B = \{0,1\}$).
The operation of the game is as follows.
The Verifier selects a pair of questions $(s,t)$ uniformly from
$\{(0,0),(0,1),(1,0)\}$ and sends $s$ and $t$ to Alice and Bob,
respectively.
Then the Verifier accepts the answers, $a$ from Alice and $b$ from Bob, if
and only if $s \vee a \neq t \vee b$.

Consider a quantum strategy for this game, where $\ket{\phi}$ is the shared
entanglement, Alice's behavior is determined by the observables $A_0$ and $A_1$,
and Bob's behavior is determined by the observables $B_0$ and $B_1$.
On input $(s,t)$, Alice computes $a$ by measuring with respect to $A_s$, and
Bob computes $b$ by measuring with respect to $B_t$.
It is straightforward to deduce that the bias of this strategy is
\begin{equation}\label{ffl-bias}
\bra{\psi}\left(
-\textstyle{\frac{1}{3}} A_0 B_0
+ \textstyle{\frac{1}{3}} A_0
+ \textstyle{\frac{1}{3}} B_0
\right)\ket{\psi}
\end{equation}
(curiously, the bias does not depend on $A_1$ or $B_1$).
Once $A_0$ and $B_0$ are determined, the optimal bias is the largest
eigenvalue of $M$, where $M = -\textstyle{\frac{1}{3}} A_0 B_0
+ \textstyle{\frac{1}{3}} A_0
+ \textstyle{\frac{1}{3}} B_0$.
Since $M^2 = -\frac{2}{3}M + \frac{1}{3}I$, this eigenvalue $\lambda$
must satisfy $\lambda^2 = -\frac{2}{3}\lambda + \frac{1}{3}$, which implies
that $\lambda = 1/3$ or $\lambda = -1$.
This implies that $\omega_q(G) \le 2/3$.
Combining this with the fact that
$2/3 = \omega_c(G \wedge G) \le \omega_q(G \wedge G) \le \omega_q(G)$,
we obtain $\omega_q(G \wedge G) = \omega_q(G) = 2/3$.

\section*{Appendix B}

In \cite{FeigeL92} it is shown that computing the classical value of
a game is equivalent to optimizing a quadratic programming
problem.
In the same paper, Feige and Lov\'asz considered two
relaxations for the quadratic programming problem. For any game $G$,
the optimum value of the first relaxation (given by Eqns.~(5)--(9) in
\cite{FeigeL92}) is denoted by $\sigma(G)$ and the optimum value of
the second relaxation (given by Eqns.~(12)--(17) in \cite{FeigeL92}) is
denoted by $\bar{\sigma}(G)$. The feasible region of the first
relaxation is subset of the feasible region of second relaxation, so
$\sigma(G)\le\bar{\sigma}(G)$. For the sake of completeness, we write
both the SDPs given in \cite{FeigeL92} for the special case of XOR games.

First, let $C$ be the matrix with entries
$C_{(s,a),(t,b)} = \pi (s,t)V(a,b|s,t)$, and let
$\hat{C}$ be the symmetric matrix
\begin{equation*}
\hat{C} = \frac{1}{2}
\left( \begin{array}{cc}
0 & C \\
C^T & 0
\end{array} \right).
\end{equation*}

\noindent The following two SDPs are relaxations of the classical value of an
XOR game, as given in \cite{FeigeL92}, with optimum value $\sigma(G)$ and
$\bar{\sigma}(G)$, respectively:
\begin{eqnarray}
\nonumber \sigma(G) &=& \max~\tr(\hat{C}P) \\
\nonumber && \text{subject to} \\
&& \sum_{a\in \01}\sum_{b\in\01} P_{(s,a),(t,b)} = 1,
\text{\ \ \ }\forall s,t \in S\cup T, \label{ModifiedConstraints} \\
&& P_{(s,a),(t,b)} \ge 0,
\text{\ \ \ }\forall s,t \in S \cup T, a,b \in \01, \label{RemovedConstraints}\\
\nonumber && P \succeq 0,
\end{eqnarray}
and
\begin{eqnarray}
\nonumber \bar{\sigma}(G) &=& \max~\tr(\hat{C}P) \\
\nonumber && \text{subject to} \\
&& \sum_{a\in \01}\sum_{b\in\01} \lvert P_{(s,a),(t,b)} \rvert \leq 1,
\text{\ \ \ }\forall s,t \in S\text{ or }s,t\in T, \label{ModifiedConstraints-1} \\
&& P_{(s,a),(t,b)} \ge 0,
\text{\ \ \  }\forall s \in S,t \in T, a,b \in \01,
\label{RemovedConstraints-1}\\
\nonumber && P \succeq 0.
\end{eqnarray}

\noindent We have the following theorem.
\begin{thm}\label{FLCSUU}
For any XOR game $G$, $\omega_q(G) = \sigma(G) = \bar{\sigma}(G)$.
\end{thm}
\begin{proof}
Let $G$ be an XOR game. From \cite{FeigeL92} we know that $\sigma(G)
\leq \bar{\sigma}(G)$. We will first show that $\omega_q(G) \leq
\sigma(G)$. For this, assume an optimal strategy for $G$. By
Theorem~\ref{Tsirel}, we can assume that $\ket{\psi}$ is a maximally
entangled state. Now let the optimal quantum strategy for $G$ be
described by the POVMs $\{M^a_s\}_{a \in \01}$, $\{N^b_t\}_{b \in
\01}$, where $s \in S$ and $t \in T$, along with the state
$\ket{\psi}$. Define
\begin{equation*}
    x_s^a = \begin{cases} (M^a_s \otimes I) \ket{\psi} & s \in S \\
                            (I \otimes N^a_s) \ket{\psi} & s \in T.
                \end{cases}
\end{equation*}
Let $\hat{P}_{(s,a),(t,b)} = x_s^a \cdot x_t^b$, so $\hat{P} \succeq 0$. It is easy to check that Eq.~\ref{ModifiedConstraints} holds,
using the fact that $M^0_s + M^1_s = I$ and $N^0_t + N^1_t = I$. For
positive semidefinite matrices like $M^a_s$ and $N^b_t$ we have that
\begin{equation*}
    \bra{\psi} M^a_s \otimes N^b_t \ket{\psi} \geq 0.
\end{equation*}
Since $\ket{\psi} = \frac{1}{\sqrt d} \sum_{k=1}^d\ket{k}\ket{k}$ is a maximally entangled state, we also have
\begin{eqnarray*}
    \bra{\psi} I \otimes N^a_s N^b_t \ket{\psi} \ = & \textstyle{\frac{1}{d}}\tr ((N^a_sN^b_t)^T) & \geq \ 0
    \\
    \bra{\psi} M^a_s M^b_t \otimes I \ket{\psi} \ = & \textstyle{\frac{1}{d}}\tr (M^a_sM^b_t) & \geq \ 0.
\end{eqnarray*}
Therefore $\hat{P}_{(s,a),(t,b)} \geq 0$ for all $s,t \in S\cup T$ and
$a,b \in \01$ and hence Eq.~\ref{RemovedConstraints} holds. With this formulation, we can turn an optimal quantum
strategy for $G$ (on a maximally entangled state $\ket{\psi}$) into
a feasible solution of (5)--(9) in \cite{FeigeL92} with same
objective value. Hence, $\omega_q(G) \leq \sigma(G)$.

Now, we will show that $\bar{\sigma}(G) \leq \omega_q(G)$.
Assume an
optimal solution $\bar{P}$ for (12)--(17) in \cite{FeigeL92}. Since,
$\bar{P}$ is a positive semidefinite matrix, we can find vectors
$x_s^a$, for $s \in S$, $a \in \01$ and $y_t^b$, for $t \in T$, $b
\in \01$, such that
\begin{equation*}
\bar{P}_{(s,a),(t,b)} = \begin{cases} x_s^a \cdot x_t^b & s,t \in S \\
                            y_s^a \cdot y_t^b & s,t \in T \\
                            x_s^a \cdot y_t^b & s \in S,t \in T.
                \end{cases}
\end{equation*}
We can view $\{x_s^a\}_{a \in \01}$ as Alice's collection of vectors
for each question $s \in S$ and $\{y_t^b\}_{b \in \01}$ as Bob's
collection of vectors for each question $t \in T$. From Eq.~\ref{ModifiedConstraints-1},
\begin{equation*}
\sum_{a,b \in \01} |x_s^a \cdot x_s^b|\leq 1
\end{equation*}
which implies
\begin{equation*} \bigg\vert \sum_{a,b \in \01}
x_s^a \cdot x_s^b\bigg\vert = \bigg\Vert \sum_{a \in
\01}x_s^a\bigg\Vert^2 \leq 1.
\end{equation*}
Therefore, $\sum_{a \in \01}x_s^a$ lie in a unit ball. By similar
argument, $\sum_{b \in \01}y_t^b$ also lie in a unit ball. Define
$x_s: = x_s^0 - x_s^1$ and $y_t: = y_t^0-y_t^1$. Now,
\begin{equation}\label{1} \bigg(\sum_{a \in \01}x_s^a\bigg) .
\bigg(\sum_{b \in \01}y_t^b\bigg) = x_s^0 \cdot y_t^0 + x_s^1 \cdot
y_t^1 + x_s^0 \cdot y_t^1 + x_s^1 \cdot y_t^0 \leq 1
\end{equation}
and we have
\begin{equation}\label{2}
x_s.y_t = x_s^0 \cdot y_t^0 + x_s^1 \cdot y_t^1 - x_s^0 \cdot y_t^1 - x_s^1 \cdot y_t^0.
\end{equation}
Therefore $x_s^0 \cdot y_t^0 + x_s^1 \cdot y_t^1 \leq  1 - (x_s^0
\cdot y_t^1 + x_s^1 \cdot y_t^0)$ and $x_s \cdot y_t \leq 1 -
2(x_s^0 \cdot y_t^1 + x_s^1 \cdot y_t^0)$, which implies
\begin{equation}\label{3}
x_s^0 \cdot y_t^1 + x_s^1 \cdot y_t^0 \leq \frac{(1 - x_s \cdot y_t)}{2}.
\end{equation}
Similarly, $x_s^0 \cdot y_t^1 + x_s^1 \cdot y_t^0 \leq 1 - (x_s^0
\cdot y_t^0 + x_s^1 \cdot y_t^1)$ and $x_s \cdot y_t \geq 2(x_s^0
\cdot y_t^0 + x_s^1 \cdot y_t^1) - 1$, which implies
\begin{equation}\label{4}
x_s^0 \cdot y_t^0 + x_s^1 \cdot y_t^1\leq \frac{(1 + x_s \cdot y_t)}{2}.
\end{equation}
{}From Eqns.~\ref{3} and \ref{4}, $\bar{\sigma}(G)$ is upper bounded by
\begin{equation*}
    \sum_{s,t} \pi(s,t) \frac{1}{2} \begin{cases} (1 + x_s \cdot y_t) &
        \text{ if the correct answer is } 0 \\
                (1 - x_s \cdot y_t) & \text{ if the correct answer is }
                1,
            \end{cases}
\end{equation*}
which is at most $\omega_q(G)$ (see Proposition 5.7 in~\cite{CleveH+04a}). Hence, $\bar{\sigma}(G) \leq \omega_q(G)$.
\end{proof}

In~\cite{FeigeL92}, it is also shown that the second relaxation is
multiplicative but the first relaxation is not. Combining our
Theorem~\ref{thm:parallel} and the multiplicativity of
$\bar{\sigma}$, we can deduce the following.
\begin{prop}
For any XOR games $G_1, \dots, G_n$,
\begin{equation*}
\omega_q(\wedge_{i=1}^n G_i) = \sigma(\wedge_{i=1}^n G_i) =
\bar{\sigma}(\wedge_{i=1}^n G_i).
\end{equation*}
\end{prop}

\end{document}